\documentclass[preprint,11pt]{elsarticle}
\usepackage[utf8]{inputenc}   
\usepackage[a4paper, total={6in, 8in}]{geometry}    
\usepackage{graphicx}    
\usepackage{amsmath}    
\usepackage{amsfonts}    
\usepackage{array}    
\usepackage{textcomp}
\usepackage{gensymb}  
\usepackage{float}    
\usepackage{multirow}   
\usepackage{url}
\usepackage{geometry}
\usepackage{array}
\usepackage{booktabs}
\usepackage[table,xcdraw]{xcolor}
\usepackage{longtable}


\newcommand{\MG}[1]{\textcolor{red}{#1}}
\begin{document}

\begin{frontmatter}

\title{A Combined Microbeam and Phase-Field Approach to Identify the Toughness and Ultimate Strength of Amorphous Silica}

\author[1]{Gustavo Alberto Rosales-Sosa\corref{cor1}}
\ead{rgustavo@neg.co.jp}
\cortext[cor1]{Corresponding author}
\author[2]{Gergely Moln{\'a}r}
\author[1]{Yoshinari Kato}
\author[1]{Taichi Nakashima}
\author[3]{Takahito Ohmura}
\author[4]{Etienne Barthel}
\author[5]{Guillaume Kermouche}
\author[5]{Sergio Sao Joao}
\author[1]{Shingo Nakane}
\author[1]{Hiroki Yamazaki}

\affiliation[1]{organization={Fundamental Technology Division, Nippon Electric Glass Co., Ltd.},
               addressline={2-7-1},
               city={Otsu},
               state={Shiga},
               postcode={520-8639},
               country={Japan}}

\affiliation[2]{organization={CNRS, INSA Lyon, LaMCoS, UMR5259},
               city={Villeurbanne},
               postcode={69621},
               country={France}}

\affiliation[3]{organization={Research Center for Structural Materials, National Institute for Materials Science (NIMS)},
              addressline={1-2-1},
              city={Tsukuba},
              state={Ibaraki}, 
              postcode={305-0047},
              country={Japan}}

\affiliation[4]{organization={Soft Matter Sciences and Engineering, ESPCI Paris, PSL University, CNRS, Sorbonne University},
               city={Paris},
               postcode={75005},
               country={France}}
\affiliation[5]{organization={Mines Saint-Etienne, Univ Lyon, CNRS, UMR 5307 LGF, Centre SMS},
               addressline={158 cours Fauriel},
               city={Saint-Etienne},
               postcode={42023},
               country={France}}

\begin{abstract}
This work presents a new approach to evaluating the toughness, described by the critical energy release rate ($G_c$), and ultimate tensile strength ($\sigma_c$) of amorphous silica (SiO$_2$ glass), combining microbeam tests and phase-field calculations. The latter provides a numerical route to brittle fracture without prescribing explicit fracture surfaces \textit{a priori}, enabling crack initiation and propagation to be tracked. Single-notched microbeams and newly designed bone-shaped microbeams with a notch-free gauge section were fabricated by Focused Ion Beam (FIB) milling nd tested under bending in air, probing the brittle-fracture and strength-controlled regimes, respectively. Both geometries were modeled by Finite Element Analysis (FEA) coupled with a phase-field formulation. We found $G_c = 5.1$~J/m$^2$ (critical stress
intensity factor $K_{IC} = 0.61$~MPa$\cdot$m$^{1/2}$), an intrinsic material length scale $\ell_c = 9.1$~nm, and $\sigma_c = 6.8$~GPa, consistent with previously reported brittle
properties of silica glass. Through a parametric study, we show the effect of notch geometry on the fracture response of the microbeams and the impact of dimensional measurement error on the determined toughness. Unlike conventional micromechanical methods that yield only $K_{IC}$, our combined microbeam geometries and phase-field approach simultaneously deliver $G_c$ and $\sigma_c$, bridging brittle-fracture characterization and the strength-controlled regime
inaccessible to toughness-only techniques.
\end{abstract}


\end{frontmatter}
\noindent

\newpage

\section{Introduction}

The mechanical reliability of silicate glasses is undermined by their extreme surface sensitivity to flaws which act as stress concentrators triggering fracture before the structural element can macroscopically reach its intrinsic strength. As a result, the usual practical strength of glass falls several orders of magnitude below the theoretical strength predicted from interatomic bonding considerations \cite{mould1959, varshneya2013, wondraczek2011}. Therefore, predicting the fracture response of glass materials under real conditions is of utmost importance for the design of damage-tolerant glass products.

Silicate glasses are archetypal isotropic brittle solids. They behave in a clearly linear-elastic manner and fail catastrophically under tensile stress although susceptible to environmentally assisted subcritical crack growth \cite{wiederhorn1967}. In the past, considerable effort has been devoted to measuring the brittle response of glass; however, very few works have directly quantified the fracture surface energy and intrinsic strength of glass. Wiederhorn's double-cantilever cleavage measurements \cite{wiederhorn1969fracture} provided an estimation of fracture surface energy $\gamma$ of a few J/m$^2$ for silicate glasses, Inagaki \textit{et al.}~\cite{inagaki1985_1} directly evaluated the work of fracture (i.e., the fracture surface energy) from Chevron-Notched Short Rod (CNSR) experiments later adapted by Lucas \textit{et al.}~\cite{lucas1995_1} for direct evaluation of $K_{IC}$. Recently, Cui and Vinci~\cite{cui2017chevron} evaluated $G_c$ from a combined FEA and bowtie-type Chevron notch microbeam approach. To \textit{et al.} \cite{to2018_1} evaluated $G_c$ from the work of fracture (fracture surface energy) using CNSR and $K_{IC}$ using the Single-Edge Precracked Beam (SEPB), respectively. More recently, Rouxel \textit{et al.} \cite{rouxel2017_1, rouxel2017_2} reported calculation of fracture surface energy of glass from semi ab-initio considerations: average single bond energy and experimental density of glass. In contrast, several works have evaluated $G_c$ of glass indirectly from the measured $K_{IC}$ values using Irwin's stress intensity factor relation \cite{lucas1995_1, smith1975fracture,  rodriguez2016_1, MATOY2009}. It is noted that  $K_{IC}$ depends strictly on the loading conditions while $G_c$ is a material property independent of the type of loading. The intrinsic strength of glass has also been evaluated in silicate glasses, mainly using fibers; however, the literature remains limited to a few works, as in the case of $G_c$. Proctor \textit{et al.}~\cite{proctor1967strength} reported the tensile strength of silica glass fibers tested by uniaxial tension in air.  Weber \textit{et al.} \cite{weber1999growth} evaluated the strength of mullite-composition glass fibers having a diameter of about 20 $\mu$m when applying pure uniaxial tension. Kurkjian and Tang \cite{kurkjian2003intrinsic, tang2015_1} reported the evaluation of ultimate strength of glass using different immersion liquids and temperatures through 2-point bending test of glass fibers. More recently, Brambilla \textit{et al.}~\cite{brambilla2009_1} and Luo \textit{et al.} \cite{luo2016_1} evaluated the tensile strength of silica glass nanofibers by pure uniaxial tension in air and using in-situ TEM, respectively. Other evaluation methods include the ring-on-ring test, which usually yields much lower strength values than fiber-based evaluations \cite{karlsson2023nondestructive, to2019} mainly due to the higher probability of finding critical surface defects in larger samples. 

Over the last decades, the advent of focused ion beam (FIB) machining and  micromechanical testing has moved these measurements to the micron scale, where flaw populations can be controlled and individual cracks sculpted. Recently, Li \textit{et al.}~\cite{li2026_1} evaluated the fracture surface energy of single-crystal SiC using the double cantilever beam (DCB) approach, while Mueller and {\v{Z}}agar \cite{mueller2015fracture, zagar2016fracture} evaluated $K_{IC}$ of amorphous silica using Chevron-notched microbeams with square and triangular cross-sections, respectively. Bruns \textit{et al.}~\cite{bruns2020fracture} estimated $K_{IC}$ in amorphous silica using a micropillar splitting approach. Harding \textit{et al.}~\cite{harding1994_1} and Scholz \textit{et al.}  \cite{scholz2004fracture} evaluated $K_{IC}$ from nanoindentation experiments in silica glass. Most of these works have focused on $K_{IC}$ but not on $G_c$ or the intrinsic strength of glass. This is mainly due to the lack of a consistent framework to numerically evaluate the experimental results, particularly for the evaluation of strength.

Due to a long-standing contradiction, neither fracture mechanics nor traditional strength-based verification can reliably predict failure. When a crack is present, the stress field becomes singular at the crack tip; thus, a strength-based criterion would predict a zero critical load. Conversely, if the sample is intact, fracture mechanics would predict an infinite critical load \cite{griffith1921,griffith1924}. In practice, however, failure occurs regardless of the validity of either criterion. A solution to this problem was introduced by Leguillon \cite{leguillon2002_1} with the coupled criterion (CC). This criterion states that a crack may initiate and propagate over a finite distance only if two criteria are met simultaneously: 1) the stress exceeds a critical value ($\sigma_c$), and 2) the incremental energy release rate exceeds a critical value ($G_c$). This method has proven effective when the initial singularity and crack path are known \textit{a priori}. A more contemporary approach is the phase-field method \cite{bourdin2000,molnar2017FEAD}, which regularizes the stress field near the crack tip and enables the prediction of failure independently of the initial geometry. The phase-field approach is rooted in Griffith’s original concept \cite{griffith1921}, which states that a crack can propagate only if sufficient elastic energy is released from the body. Griffith postulated that creating a unit crack surface requires an energy expenditure of $G_c$, the critical energy release rate, an intrinsic material parameter. The phase-field formulation also introduces a second material constant ($\ell_c$) that governs the extent of damage and, consequently, the degree of stress-field regularization. Recent works demonstrated a direct and geometry-independent correlation~\cite{molnar2020TAFMAT,molnar2022EFM} between $\ell_c$ and $\sigma_c$.

To achieve the evaluation of both $G_c$ and $\ell_c$, and ultimately $\sigma_c$, it is therefore necessary to design experimental setups which allow for independent calibrations and isolate the brittle-fracture from the strength-controlled regimes, combined with a numerical framework that does not presuppose an explicit crack path.  In the present study, we design two complementary micron-scale geometries---a single-notched geometry and a newly designed bone-shaped geometry---obtained by FIB and tested under bending. These experiments allowed us to calibrate $G_c$ and $\ell_c$ of amorphous silica, which gave access to the evaluation of $\sigma_c$. The geometries were modeled by finite element analysis coupled with a brittle-fracture phase-field formulation, which removes the need to prescribe the fracture surface \textit{a priori}. In this work, the brittle-ductile phase-field formulation~\cite{molnar2020CMAME} was not used because we did not observe any plasticity in tension. The paper is structured as follows: Section~\ref{sec:methods} introduces the phase-field formulation and its two governing material parameters, describes the microbeam geometries, their FIB fabrication, and the bending test protocol. Section~\ref{sec:results} presents the measured load–displacement responses together with the coupled FEA–phase-field analysis used to extract $G_c$, $K_{IC}$, $\ell_c$, and $\sigma_c$. Section~\ref{sec:discussion} discusses: a) the effect of crack geometry of notched microbeams on the fracture response, b) a parametric sensitivity study of the geometry on the evaluated $G_c$ and stiffness, and lastly c) the connection between glass atomic structure and phase-field parameters. 

\newpage

\section{Methods}\label{sec:methods}


In this section, we describe the experimental and numerical methods employed in this study. The experimental methods primarily focus on beam fabrication and testing, while the numerical methodology briefly outlines the well-known phase-field framework. Finally, we present the principal strategy for identifying phase-field fracture properties using the novel geometry proposed in this work.


\subsection{Microbeam fabrication}

\begin{figure}[H]
    \centering
    \includegraphics[width=1.0\columnwidth]{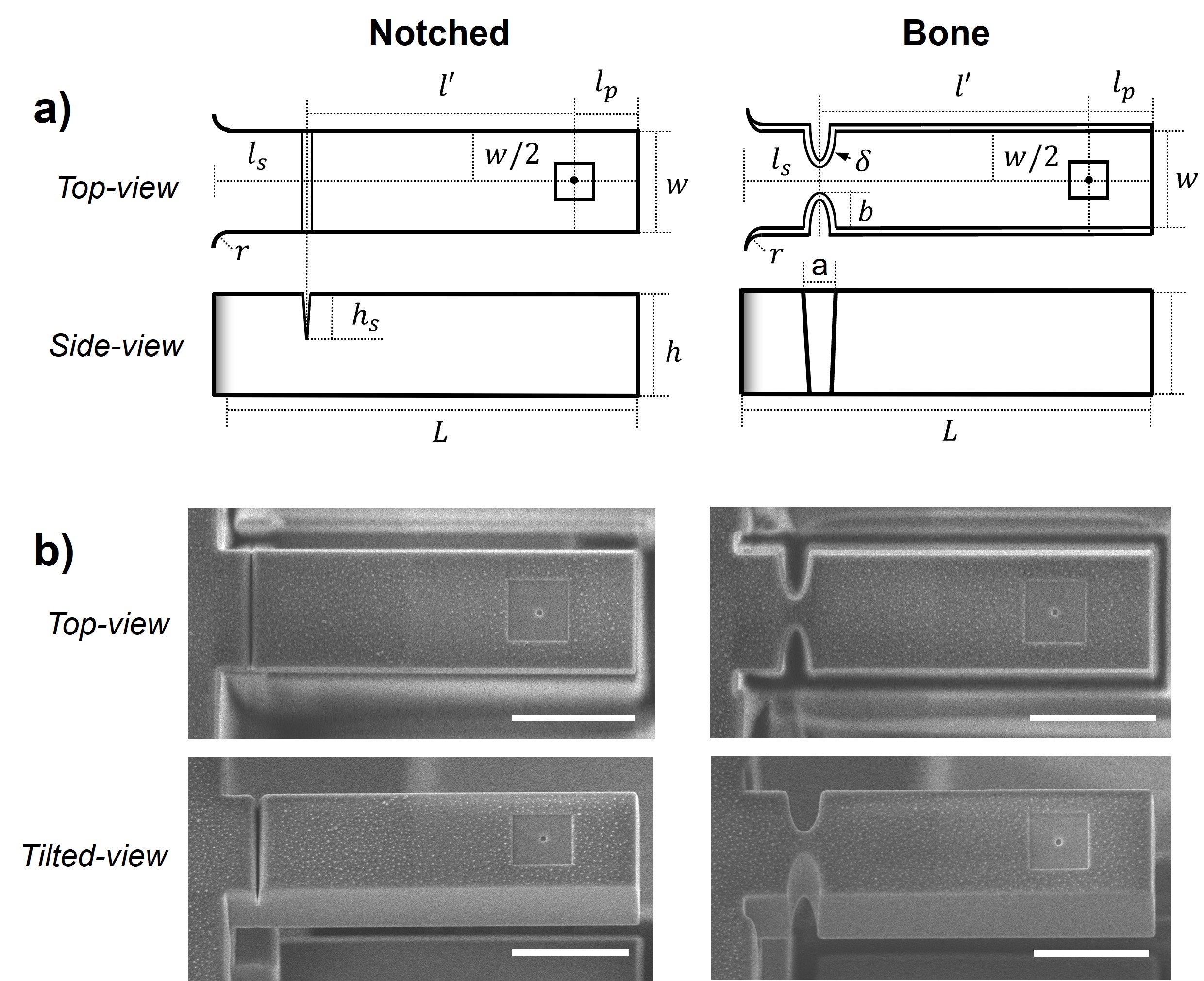}
    \caption{ a) Top- and side-view diagrams of the notched and  bone microbeam geometries b) scanning electron microscope (SEM) images of the fabricated microbeams from both top and tilted observations. The white reference mark in the images corresponds to a length of 2 $\mu$m. }
    \label{fig:geometry}
\end{figure}

Silica glass (GE-124, $<$10 ppm OH) microbeams were fabricated through FIB process (HITACHI FB2200, Japan) from Au-coated (10 nm) thin plates (5 mm x 5 mm x 0.5 mm) using an acceleration voltage of 30--40 kV and a beam aperture of 15--300 $\mu$m. The milling process started from one of the sample edges.
Two types of beam geometries were fabricated: a single-notched and a new bone-shaped geometry. Figure \ref{fig:geometry} shows the schematics of the microbeams with associated geometric parameters. The summary of geometrical parameters as measured from the SEM images is shown in Table \ref{table:parameters}. The geometry was measured from the SEM photos using image processing.

The fabrication process was carried out in several steps, starting from rough milling (large aperture) and progressing to very fine milling (small aperture) for final tuning of the geometry. During the rough milling stage, significant re-deposition of material was observed on some of the microbeam faces, which required an intermediate milling step to remove the re-deposited zones.

For the notched microbeams, the notch had to be fabricated in multiple stages. First, the notch was milled from the top on a sufficiently thick cantilever, and then the outer faces were milled to eliminate the notch deflection caused by FIB milling near the microbeam lateral surfaces \cite{brinckmann2017ce}. The notch was fabricated at a voltage of 30 kV to create a wider top opening, allowing the milled material to leave the process zone more easily. The final tuning of the geometry for all microbeams was done using a small aperture size (e.g., 15 $\mu$m) and a high voltage (40 kV). This allowed us to obtain microbeams with almost straight faces and reduced the amount of re-deposition.

\begin{table}[H]
\centering
\caption{Geometrical parameters for the notched (N) and bone (B) microbeams. All units are shown in micrometers. Parameter \textit{D} corresponds to the notch tip radius in the notched microbeams. All the parameters in the table correspond to those shown in Figure 1A. The experimental uncertainty is $\pm$0.01 $\mu$m for all dimensional parameters except $D$, for which the uncertainty was $\pm$0.001 $\mu$m.}
\vspace{0.5em}
\begin{tabular}{ccccccccccccc}
\toprule
 & $L$ & $w$ & $h$ & $l'$ & $l_s$ & $h_s$ & $l_p$ & $D$ & $\delta$ & $a$ & $b$ & $r$ \\
\midrule
Sample & \multicolumn{12}{c}{($\mu$m)} \\
\midrule
N-1 & 7.16 & 2.01 & 2.04 & 4.77 & 0.77 & 0.52 & 1.62 & 0.008 & - & - & - & 0.15 \\
N-2 & 7.15 & 2.04 & 2.05 & 4.89 & 0.63 & 0.54 & 1.63 & 0.008 & - & - & - & 0.14 \\
N-3 & 7.19 & 2.09 & 2.03 & 4.83 & 0.71 & 0.60 & 1.65 & 0.007 & - & - & - & 0.13 \\
N-4 & 7.22 & 2.05 & 2.00 & 4.91 & 0.71 & 0.58 & 1.60 & 0.007 & - & - & - & 0.13 \\
N-5 & 7.12 & 2.16 & 2.03 & 4.61 & 0.88 & 0.64 & 1.63 & 0.007 & - & - & - & 0.10    \\
N-6 & 7.09 & 2.02 & 2.00 & 4.60 & 0.89 & 0.70 & 1.60 & 0.007 & - & - & - & 0.15 \\
N-7 & 7.13 & 2.02 & 2.04 & 4.81 & 0.74 & 0.73 & 1.58 & 0.007 & - & - & - & 0.18 \\
N-8 & 7.21 & 2.11 & 2.05 & 4.86 & 0.74 & 1.02 & 1.61 & 0.007 & - & - & - & 0.15 \\
N-9 & 7.20 & 2.10 & 2.03 & 4.86 & 0.72 & 1.00 & 1.62 & 0.007 & - & - & - & 0.14 \\
\midrule
B-1 & 7.21 & 1.92 & 2.11 & 4.40 & 1.15 & -    & 1.66 & -     & 0.10  & 0.74 & 0.75 & 0.18 \\
B-2 & 7.17 & 1.93 & 2.13 & 4.42 & 1.00 & -    & 1.65 & -     & 0.10  & 0.74 & 0.74 & 0.17 \\
B-3 & 7.12 & 1.94 & 2.03 & 4.42 & 1.06 & -    & 1.64 & -     & 0.09  & 0.65 & 0.78 & 0.09 \\
B-4 & 7.14 & 1.96 & 2.10 & 4.47 & 1.05 & -    & 1.62 & -     & 0.09  & 0.67 & 0.77 & 0.08 \\
B-5 & 7.14 & 1.94 & 2.01 & 4.44 & 1.05 & -    & 1.65 & -     & 0.07& 0.68 & 0.74 & 0.11 \\
\bottomrule
\end{tabular}
\label{table:parameters}
\end{table}

\subsection{Experimental loading}

The load-displacement curves and fracture forces of the microbeams were measured in air (50\% RH) using a nanoindenter fitted with a Berkovich indenter (Hysitron-Bruker Triboindenter TI950, USA). The indenter also serves as the tip for the Scanning Probe Microscope (SPM) integrated into the nanoindenter. The SPM allows detailed scanning of the sample surface at the nanoscale and precise positioning of the load. The TI950 indenter is an inherently force-controlled system operating through a capacitive transducer with a feedback loop  (up to 78 kHz). However, all experiments were performed in displacement control mode. This particular mode is operated via a software-based feedback loop that uses real-time displacement data from the capacitive transducer to regulate the actuator's motion. However, upon sudden brittle failure, this experimental setup usually produces artificial displacement overshoots to compensate for the decrease in reaction forces. Glass plates were attached to thick steel disks with glue and left to cure overnight. A pre-load cycle to a maximum displacement of 50 nm was applied before the actual test to ensure appropriate contact. During the test, a maximum displacement of 500-1000 nm was applied at a loading rate of 5 nm/s and an acquisition rate of 600 points/s. The resulting load-displacement curves were compliance and thermal-drift corrected via software.\\

\subsection{Phase-field fracture method}

The phase-field technique \cite{bourdin2000} is based on the reformulation of the original Griffith \cite{griffith1921,griffith1924} theory on a variational basis \cite{francfort1998}. The phase-field fracture model is based on a diffuse representation of localized discontinuities, where the crack surface is approximated by a continuous damage variable $d$ ranging from $0$ to $1$. When $d = 0$, the material is fully intact, whereas $d = 1$ corresponds to complete failure, with total loss of stiffness and load-bearing capacity. This formulation enables a smooth transition from the undamaged to the fully fractured state, allowing the simulation of crack initiation and propagation without the need for explicit crack tracking. To calculate the new auxiliary variable, the total internal energy of the solid is minimized:

\begin{equation}
\Pi({\bf u}, d) = E + W = \int_\Omega \left[ g(d)\psi_0^+({\bf u}) + \psi_0^-({\bf u}) \right] \mathrm{d}\Omega
+ \int_\Omega \frac{3G_{c,pf}}{8\ell_c} \left[ d + \ell_c^2 \left| \nabla d \right|^2 \right] \mathrm{d}\Omega
\end{equation}

Here, $E$ and $W$ represent the strain and fracture energy components, respectively, and ${\bf u}$ denotes the displacement field defined over the domain $\Omega$. The fracture energy is characterized by two key parameters: the phase-field critical energy release rate $G_{c,pf}$ and the crack surface density function, while the damage diffusion is governed by $\ell_c$ also displaying the spatial gradient of the damage $\nabla$$d$ to treat singularities in a thermodynamically consistent manner \cite{molnar2022EFM}. The undamaged strain energy density $\psi_0$ is decomposed into tensile ($\psi_0^+$) and compressive ($\psi_0^-$) parts, with only the tensile component subject to degradation through a quadratic function $g(d) = (1 - d)^2$. Consequently, only tensile energies contribute to crack evolution. To distinguish between tensile and compressive strain energies, we adopt the spectral decomposition proposed by Bernard \textit{et al.} \cite{bernard2012}. Finally, the total fracture surface $\Gamma$ in the phase-field formulation is obtained by integrating the crack surface density over the domain:

\begin{equation}
\label{eq:S}
\Gamma = \int_\Omega \frac{3}{8\ell_c} \left[ d + \ell_c^2 \left| \nabla d \right|^2 \right] \mathrm{d}\Omega .
\end{equation}

The mechanical and phase-field problems were solved using a staggered time-integration. To ensure irreversibility and enforce positive damage evolution, Lagrange multipliers were implemented following the approach of Lu \textit{et al.} \cite{Lu2020}. All calculations used adaptive time stepping. Since our experiments revealed strain below 5\%, a small strain formulation (small displacement and rotation) was adopted. For comprehensive implementation details, readers are referred to our recent work~\cite{molnar2022EFM}.

\subsection{Correction of $G_c$  based on FEA mesh}
It has been shown that brittle fracture phase-field calculations are subject to numerical errors. These numerical errors originate from: 1) time discretization, 2) spatial discretization. The first error has been minimized through the staggered phase-field algorithm as proposed in the literature \cite{molnar2020TAFMAT,molnar2020CMAME}. The spatial discretization can be divided into two types: i) approximation error and ii) localization error. The first arises from the difference between the theoretical crack surface analytical solution (AT1 vs. AT2) and the FEA numerical approximation. The second arises from the crack opening at the damaged state and known to be significantly larger than the approximation error. The spatial discretization error needs to be accounted for in the evaluation of the real material critical energy release rate $G_c$ though the following relation,

\begin{equation}
  G_c =  G_{c,pf}\left(1+\alpha \frac{m_s}{c_{\omega}\ell_c}\right)
  \label{eqn:correction}
\end{equation}
where $m_s$ is the size of the FEA mesh at the crack surface, and $c_{\omega}$ is a constant that depends on the damage model type (AT1 or AT2) \cite{molnar2022EFM}. The parameter $\alpha$ describes the boundary conditions at the crack surface. In this work we used $\alpha=1$ and $c_{\omega}=8/3$, which correspond to the AT1 model without any boundary conditions. The AT1 model is consistent with an elastic threshold before failure as observed in linear elastic brittle solids \cite{pham2011_1}. The FEA solid model was designed so that the ratio $m_s/\ell_c=0.5$ (mesh size of one half the length-scale parameter), which ensures an approximation error of about 1\% and a localization error of 18.1\%. In this manuscript we will refer to the corrected critical energy release rate simply as $G_c$. 

\subsection{Model and material parameter identification}


The parameters $G_c$ and $\ell_c$ were evaluated through a combination of two experiments: notched and bone microbeam tests, allowing both the brittle-fracture-dominated (notched) and strength-controlled (bone) fracture regimes to be probed.

Our current phase-field model requires the material's $G_c$ and $\ell_c$, and elastic properties $E$ and $\nu$ as input parameters for the calculations. For the case of  notched microbeams, the singularity is very sharp (less than 10 nm), so $G_c$ can be extracted directly by fitting it to reproduce the experimental microbeam fracture forces. In order to confirm that $\ell_c$ does not have any effect on the value of $G_c$, we fitted $G_c$ using several $\ell_c$ values which were consistent with the experimental fracture forces. We will show that in sharp notched beams the identified value of $G_c$ is insensitive to $\ell_c$ as long as the crack tip radius is smaller than $\ell_c$.

For the bone microbeams, the singularities are blunt and much larger than the material's expected $\ell_c$; therefore,  the value of $G_c$ which reproduces the fracture force will depend strongly on $\ell_c$. Similarly to the notch microbeam experiments, we calculate all the possible combinations of $G_c$ and $\ell_c$ values which are consistent with the bone microbeam fracture forces. 

Finally, in the $G_c$--$\ell_c$ space, the intersection of the experimentally consistent regions for both sample geometries defines the most likely ranges of phase-field parameters for the material.

It is noted that it is also possible to evaluate the stress intensity factor from purely FEA elastic calculations in notched microbeams. In this case, we need to extract the nodal stresses or displacements as a function of distance from the notch tip, which are related to the mode I stress intensity factor as pointed out by Williams \cite{Williams} and further tested by Fujita \textit{et al.} \cite{fujita2024} in ion-exchanged glass microbeams. However, this approach does not provide access to the ultimate tensile strength, which is an advantage of our current implementation of the phase-field method.

\subsection{Evaluation of mode I stress intensity factor $K_{IC}$}

From $G_c$ it was possible to directly quantify $K_{IC}$ for the case of single-notched microbeams using the plane strain assumption $K_{IC}=\sqrt{G_c E'}$, where $E'=E/(1-\nu^2)$ is the effective modulus. The value of $K_{IC}$ was also obtained for the notched beams using the relation proposed by Matoy \textit{et al.} \cite{MATOY2009} for single-notched microbeams,

\begin{equation}
  K_{IC} = \frac{P_c l'}{w h^{3/2}} f\left(\frac{h_s}{h}\right)
  \label{eqn:Matoy}
\end{equation}

\begin{equation}
  f\left(\frac{h_s}{h}\right) = 1.46 + 24.36\left(\frac{h_s}{h}\right)-47.21\left(\frac{h_s}{h}\right)^2+75.18\left(\frac{h_s}{h}\right)^3
  \label{eqn:geom_func}
\end{equation}
where $P_c$ is the experimental fracture force of the notched microbeams. Equation \ref{eqn:Matoy} is valid for $l_s$:$l'$:$h$:$w$ = 2:5:1.1:1.7 and for $h_s/h$ ratios between 0.05 and 0.45. 

\subsection{Ultimate tensile strength evaluation}

The ultimate tensile strength is defined as the maximum pure tensile stress that the material can withstand.  The homogeneous phase-field solution \cite{molnar2022EFM,molnar2025_1}  provides the tensile strength $\sigma_c$ of the material directly from the material parameters: Young's modulus ($E$), Poisson's ratio ($\nu$), critical energy release rate ($G_c$), and the length-scale parameter $\ell_c$,

\begin{equation}
  \sigma_c = \eta \sqrt{\frac{{E G_c}}{\ell_c}}
  \label{eqn:strength}
\end{equation}
with function $\eta$ depending on the Poisson's ratio and the stress state (principal stress ratios). For the particular case of pure uniaxial tension, the principal stress ratios $\sigma_2/\sigma_1$ and $\sigma_3/\sigma_1$ equal zero. Using the determined $G_c$ and $\ell_c$ we computed the ultimate tensile strength of silica glass. Details of the numerical implementation can be found in a previous work \cite{molnar2022EFM}.

\newpage

\section{Results}\label{sec:results}

\subsection{Experimental results}
In this section we present the experimental fracture responses of the silica microbeams including load-displacement curves, fracture forces, and fractography.

\subsubsection{Load-displacement curves}

\begin{figure}[H]
    \centering
    \includegraphics[width=1.0\columnwidth]{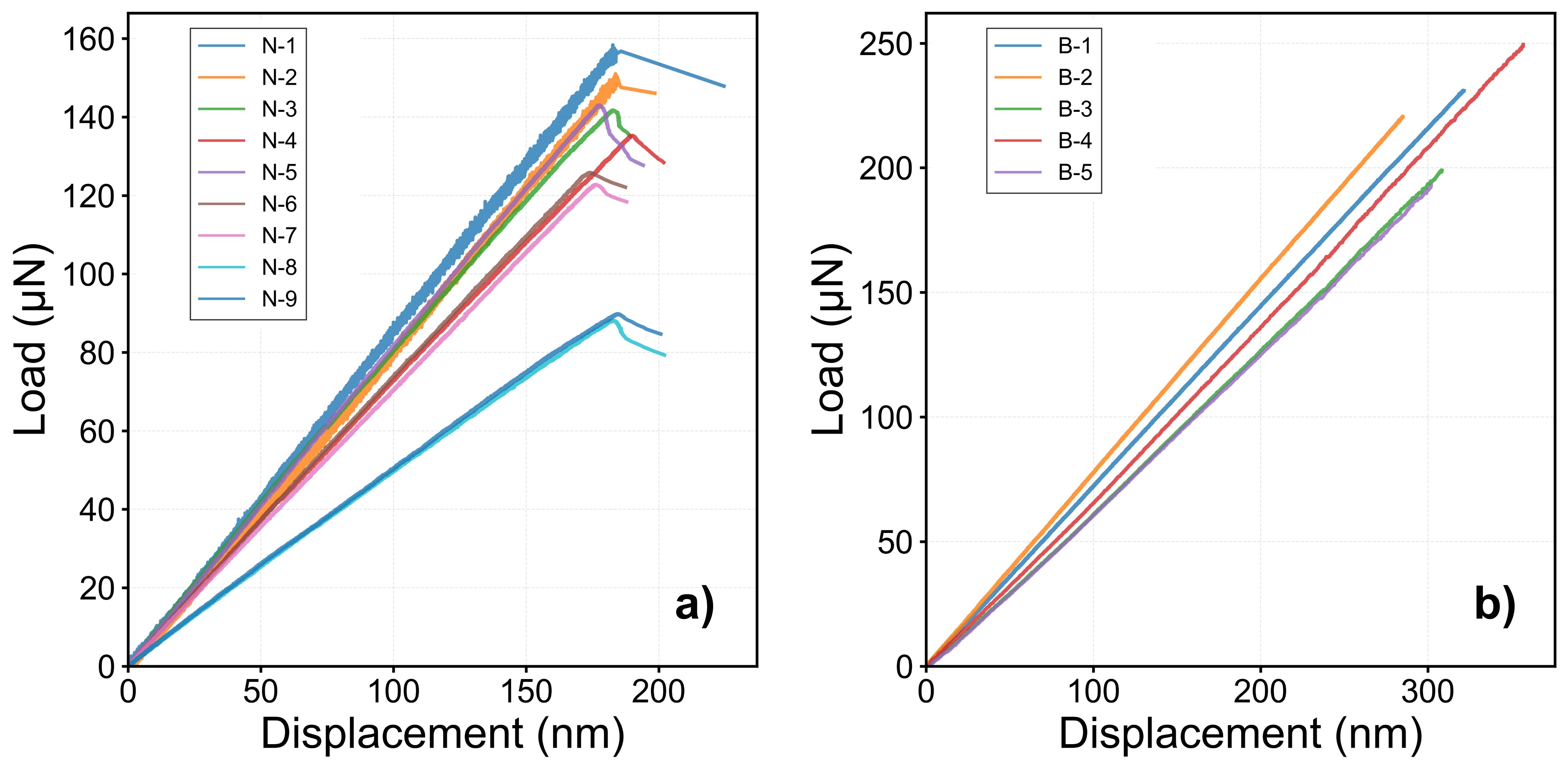}
    \caption{ Experimental load-displacement curves for a) notched and b) bone microbeams under bending test. Maximum load for each curve corresponds to $P_c$.}
    \label{fig:p_h_all}
\end{figure}

The experimental load-displacement curves for all samples are shown in Figure \ref{fig:p_h_all}. For the notched microbeams, the fracture forces and elastic stiffness decrease with increasing notch depth $h_s$. For the bone-type microbeams, the fracture forces vary slightly among the samples (see details in Table \ref{tab:fracture_force}). Overall, the fracture forces of the bone microbeams are about two times those of the notched microbeams.

All of the beams failed catastrophically after reaching a fracture load ($P_c$) and no subcritical crack growth was observed. The transducer of the indenter used in this work is inherently force-controlled; displacement is regulated via a software feedback loop and, therefore, cannot follow the displacement when the crack proceeds.

\begin{table}[H]
\centering
\caption{Fracture forces of investigated microbeam samples: Notched (N) and Bone (B).}
\vspace{0.5em}
\label{tab:fracture_force}
\begin{tabular}{cc}
\hline
Sample & Fracture force $P_\mathrm{c}$ ($\mu$N) \\
\hline
N-1 & 158.30 \\
N-2 & 151.00 \\
N-3 & 143.05 \\
N-4 & 141.73 \\
N-5 & 135.27 \\
N-6 & 125.84 \\
N-7 & 122.74 \\
N-8 & 88.14 \\
N-9 & 89.82 \\
B-1 & 231.00 \\
B-2 & 220.00 \\
B-3 & 200.70 \\
B-4 & 251.22 \\
B-5 & 195.65 \\
\hline
\end{tabular}
\end{table}

\subsubsection{Fractography}

\begin{figure}[htb]
    \centering
    \includegraphics[width=0.8\columnwidth]{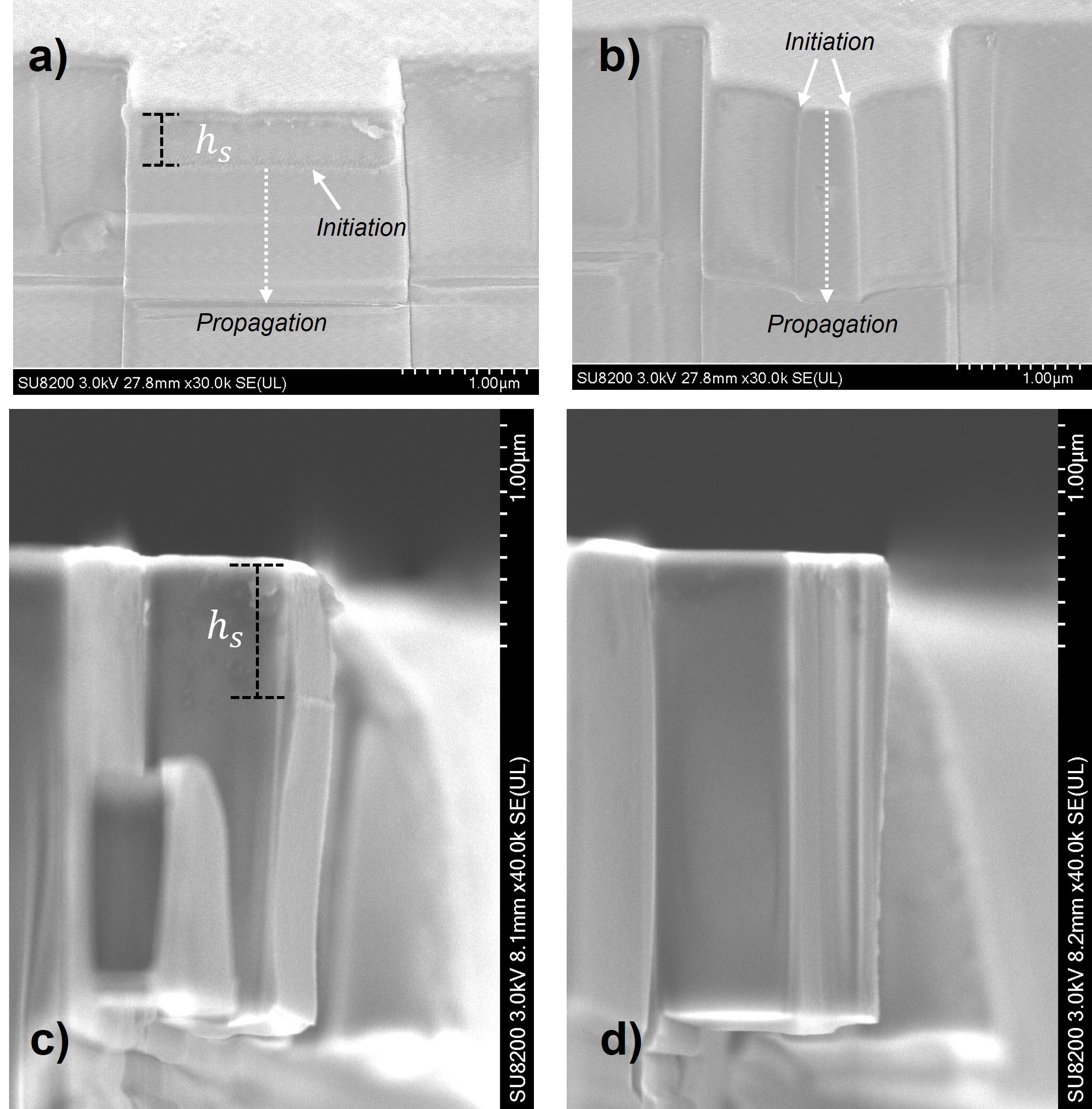}
    \caption{ Example of scanning electron microscope (SEM) images of the broken notched (a and c) and bone (b and d) microbeams showing the fracture surfaces (black dashed lines). Also indicated are the notch depth $h_s$ (frontal and side-tilted views), crack initiation zones (solid arrows) and crack propagation directions (dotted arrows). Notched and bone samples are N-6 and B-1, respectively.}
    \label{fig:cross_sections}
\end{figure}

Figure \ref{fig:cross_sections} shows an example of the scanning electron microscope observation of the microbeams after fracture. The notched microbeams show clear contrast between the fractured part and the initial notch, which allows the notch depth $h_s$ to be determined more precisely. For the case of the bone beams, the cracks initiated at the narrow section of the beam. From the side-tilted observation, the cracks seem to have propagated at a slight inward angle from the initiation point. 

\subsection{Numerical results}
In this section we show the numerical results of the investigation: FEA model, damage evolution, and material parameter identification.

\subsubsection{FEA model}
\begin{figure}[H]
    \centering
    \includegraphics[width=1.0\columnwidth]{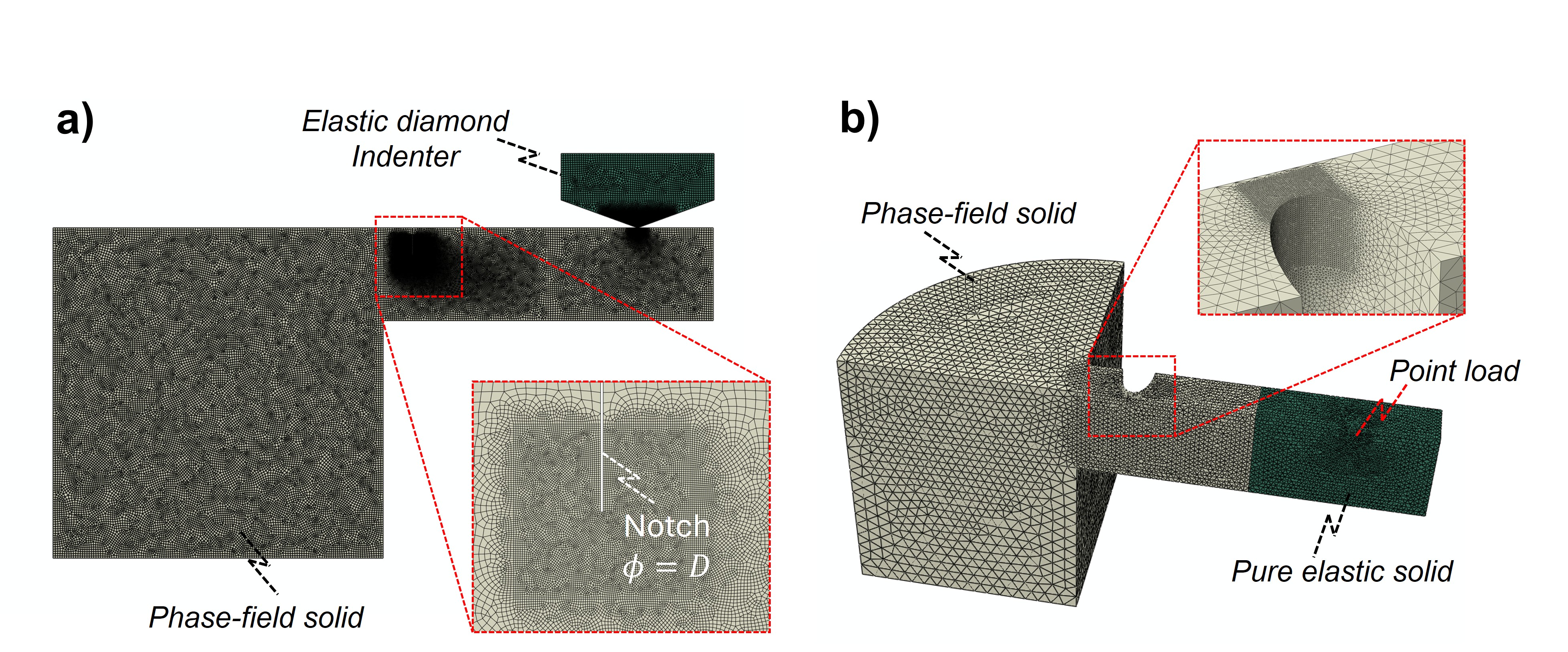}
    \caption{ Example of FEA models used for the phase-field calculations using an $\ell_c$ = 20 nm: a) Notched microbeams (2D - plane strain) showing the full model and zoom on notched area, and b) bone microbeams (3D - 1/2 symmetry) with zoom in the singularity. The green (dark) areas are modeled as fully elastic while the gray areas are elastic-brittle phase-field elements.}
    \label{fig:models}
\end{figure}

FEA simulations were performed in ABAQUS 2022 FEA software. FEA solid models were built for both notched and bone geometries. For the case of notched microbeams (Figure \ref{fig:models}A), we used plane strain 2D models (quad CPE4 elements). For bone microbeams (Figure \ref{fig:models}B), we developed half-symmetry 3D models (4-noded C3D4 tetra-elements). The mesh at the core region where the stress concentrator is located, was about 10 times the size of $\ell_c$ to ensure numerical accuracy of the damage variable $d$.  ABAQUS Python scripts that adapt the core area mesh to the value of $\ell_c$ were written for each microbeam geometry. Bending simulations were performed using displacement control at the loading point  of the microbeams using a maximum displacement of 1 $\mu$m. The damage field ($d$)  was computed through a combined user-defined element (UEL) and user-defined material subroutine (UMAT). The numerical implementation of the subroutines is shown in detail in a previous work \cite{molnar2022EFM}.  For the simulations, the elastic properties of silica ($E = 72$ GPa and $\nu = 0.17$) and diamond ($E = 1140$ GPa and $\nu = 0.07$) shown in our previous work \cite{ROSALES2025} were used. The diamond indenter and the end part of the bone-microbeam apart from the neck region (Figure \ref{fig:models}b) were modeled as fully elastic to avoid phase-field convergence issues due to sharp contact nonlinearities.  

\newpage

\subsubsection{Damage evolution}
\begin{figure}[H]
    \centering
    \includegraphics[width=0.8\columnwidth]{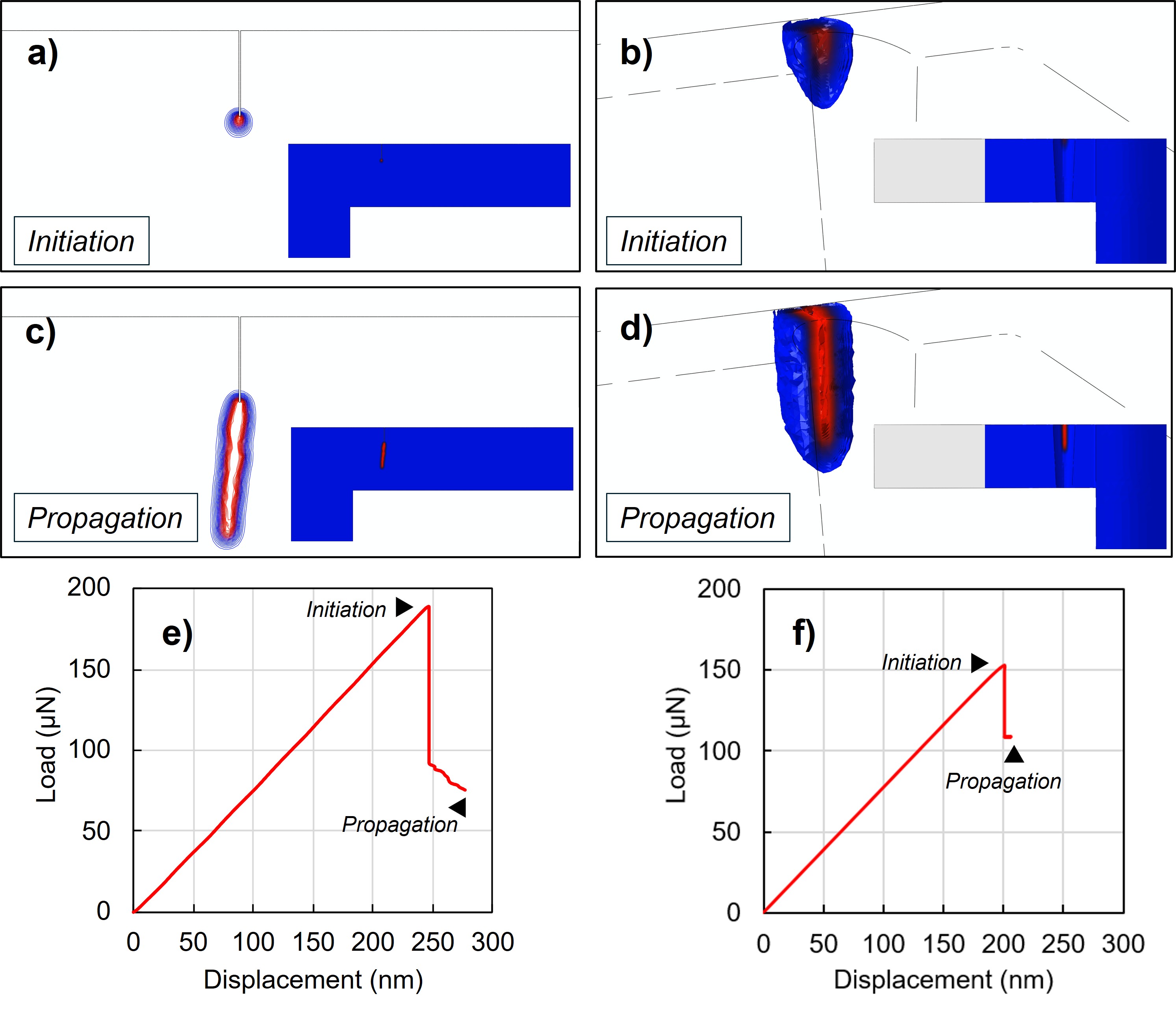}
    \caption{Phase-field damage maps at the point of fracture initiation and after propagation along with their respective load-displacement diagrams (bottom): (a,c,e) notched microbeams, and (b,d,f) bone microbeams. The damage field $d$ varies between blue (no fracture, $d=0$) and red (fully fractured, $d=1$). Insets show the damage field in the whole model. For this example in both cases $G_c = 6$ J/m$^2$ with a length-scale parameter of $\ell_c = 50$ nm and $\ell_c = 100$ nm for the notched and bone geometries, respectively.}
    \label{fig:propagation}
\end{figure}

One of the main advantages of the phase-field approach \cite{molnar2022EFM} is its ability to model crack initiation and propagation in a thermodynamically consistent manner, based solely on the competition between the elastic strain energy and fracture energy dissipation, without requiring explicit crack tracking or predefined crack path. To illustrate this point, Figure \ref{fig:propagation} shows calculation examples for the cases of crack initiation and propagation for both beam geometries along with their load-displacement curves. The results show the damage field at initiation and after propagation.
 For the case of the notched beam, the crack initiates and propagates off-axis at a slight tilt towards the base of the cantilever. This was indeed observed in the experiments (see Figure \ref{fig:cross_sections}). The slight tilt is due to the fact that the stress field has a slight shear component as shown previously by Molnár \textit{et al.} \cite{molnar2017FEAD} when testing fracture under pure uniaxial tension and pure shear cases. For the case of the bone beam, the fracture initiates at the edge near the singularity and propagates downwards and inwards simultaneously. In both cases the force decreases sharply upon fracture.  

\subsubsection{Material parameter identification}

\begin{figure}[H]
    \centering
    \includegraphics[width=0.8\columnwidth]{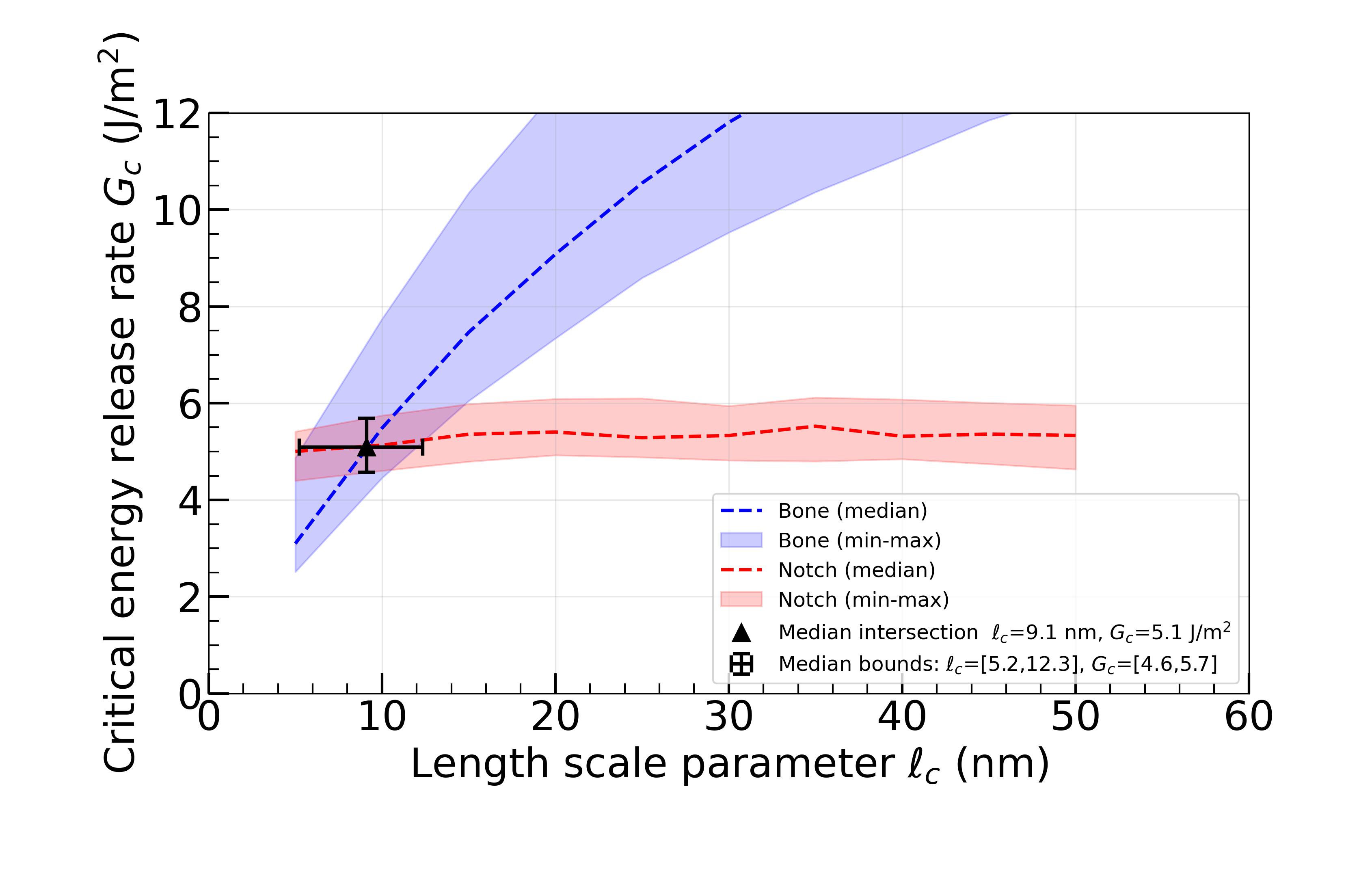}
    \caption{ Simulated fracture force-consistent $G_{c}$ and $\ell_c$ for the notched (light red) and bone (light blue) microbeams. The optimal values correspond to the intersection of the median curves (dashed lines) and variation is given by the bounds (black lines). The values of intersection and its respective bounds for all the samples are shown in the legend.}
    \label{fig:optimal_values}
\end{figure}

The simulated fracture-force-consistent combinations of $G_c$ and $\ell_c$ values for both notched and bone microbeam experiments are shown in Figure \ref{fig:optimal_values}. The figure shows the experimental variation bands and the median curve in each case. For the notched beam, the $G_c$ values look constant even at different $\ell_c$, while for the bone beam the value of $G_c$ increases with the $\ell_c$ value. The intersection of the curves provides the optimal material parameters $G_{c}$ and $\ell_c$, with bounds defined by the black bars accounting for the experimental variation of both parameters. 

From $G_c$ we computed $K_{IC}$ using the plane strain approximation. Lastly, from $\ell_c$, we computed the ultimate tensile strength following equation \ref{eqn:strength}. Table \ref{tab:fracture_params} and table \ref{tab:fracture_lit} show the currently obtained results along with other reported data in the literature for silica glass. The estimated values of $G_c$, $\ell_c$, $K_{IC}$, and $\sigma_c$ for silica were 5.1 J/m$^2$, 9.1 nm, 0.61 MPa$\cdot$m$^{1/2}$ and 6.8 GPa, respectively. For the case of $G_c$, Wiederhorn reported a value of 8.7 J/m$^2$ for silica glass when tested in liquid $N_2$ and Cui and Vinci suggested a $G_c$ value of 5.2 J/m$^2$ for silica glass when tested in vacuum. The theoretical model by Rouxel \cite{rouxel2017_1, rouxel2017_2} predicts a $G_c$ of 7.3 J/m$^2$ for silica glass when using the single bond strength approximation. Overall, our results are within the literature range. 

Regarding the length-scale parameter $\ell_c$, previous literature shows smaller values compared to our results; however, those were estimated from theoretical considerations \cite{tomic2022phase} or molecular dynamics simulations \cite{molnar2024_1}; to our knowledge, the evaluated $\ell_c$ values are the first reported for silica glass obtained from experiments.

The obtained values of $K_{IC}$ for silica glass are within the range of literature values. We note that macroscopic experiments such as the SEPB method give consistently larger values, by about 10--20\%, than those obtained from micromechanical methods such as the chevron-notched microbeam test \cite{mueller2015fracture, zagar2016fracture}, the micropillar splitting method \cite{bruns2020fracture}, or the single-notched beam (current study). The $K_{IC}$ values found through the analytical solution (equation \ref{eqn:Matoy}) are larger than those obtained by our method, closer to those reported by macroscopic tests \cite{lucas1995_1,smith1975fracture} and similar to that reported by Liu \textit{et al.}~\cite{liu2023_1} for silica microbeams using the same equation. It is noted that classical evaluations of $K_{IC}$ and $G_c$ using Linear Elastic Fracture Mechanics (LEFM) idealize the crack as infinitely sharp; however, in reality cracks may have finite radius which might lead to an overestimation of the fracture surface energy inflating both $K_{IC}$ and back-calculated $G_c$ values. As the phase-field method regularizes the crack tip over the intrinsic length scale, it is expected to provide a closer representation to the fracture phenomena. 

The evaluated strength values match closely the reported data for silica glass in air. The strength values reported by Tang \textit{et al.}~\cite{tang2015_1} in liquid nitrogen and by Luo \textit{et al.} \cite{luo2016_1} in vacuum, are about two times larger than that in air. This suggests that humidity may have a significant impact on the intrinsic strength of silica glass either through reduction of $G_c$ or increasing $\ell_c$ at the surface. These strength values are expected to be larger when tested in inert environments possibly due to a ``less water-modified'' atomic structure at the surfaces. It remains for future studies to evaluate the strength of glass in vacuum.

Finally, it is noted that Ga$^+$ ions implanted during the FIB process can affect the fracture response of the material. Usually conventional FIB process lead to some implantation of Ga$^+$ ion over a few nanometers around fabricated notches using ``overFIBing'' method \cite{brinckmann2017ce}; yet our obtained values are consistent with the reported literature value indicating that Ga$^+$ effect on the mechanical properties remains limited. Nevertheless, a systematic assessment of Ga$^+$ ion-damage effects remains a worthwhile direction for future investigations.

\begin{table}[H]
\centering
\caption{Brittle fracture parameters for amorphous silica evaluated in air from notched and bone microbeam tests (this work), with the overall range of literature values (see Table~\ref{tab:fracture_lit}). Individual literature sources are listed in Table~\ref{tab:fracture_lit}. Values of $K_{IC}$ in parentheses were obtained from the analytical solution (Eq.~\ref{eqn:Matoy}); all others from phase-field calculations. The literature range includes results for tests in air and in vacuum.}
\vspace{4pt}
\label{tab:fracture_params}
\begin{tabular}{lccccc}
\toprule
Parameter & Unit & Median & Min. & Max. & Literature range \\
\midrule
$G_c$      & J/m$^2$            & 5.1         & 4.6         & 5.7         & 5.2--8.7 \\
$\ell_c$   & nm                 & 9.1         & 5.2         & 12.3        & 2.0--3.0 \\
$K_{IC}$   & MPa$\cdot$m$^{1/2}$ & 0.61 (0.75) & 0.58 (0.72) & 0.64 (0.78) & 0.58--0.78 \\
$\sigma_c$ & GPa                & 6.8         & 5.3         & 9.1         & 5.1--13.0 \\
\bottomrule
\end{tabular}
\end{table}

\begin{table}[H]
\centering
\caption{Reference data for amorphous silica brittle fracture parameters, with testing condition and method. DCC -- double-cantilever cleavage;  CNMB -- chevron-notched microbeam; PF -- phase-field; MD -- molecular dynamics; NB -- notched-beam; CNSR -- chevron-notched short rod; NI -- nanoindentation; PS -- pillar splitting;  NMB -- notched microbeam; 2PB -- two-point bending (fibers); ROR -- ring-on-ring. {$\dagger$} Back-calculated from the reported $K_{IC}=0.62$ MPa$\cdot$m$^{1/2}$ via $G_c = K_{IC}^2(1-\nu^2)/E$ ($E=72$ GPa, $\nu=0.17$); the original paper reports $K_{IC}$, not $G_c$. {$\ddagger$} Values measured on nanofibers with diameters below 5 nm.}
\vspace{4pt}
\label{tab:fracture_lit}
\setlength{\tabcolsep}{6pt}
\begin{tabular}{llllll}
\toprule
Parameter & Value & Condition & Method & Author & Ref. \\
\midrule
\multirow{3}{*}{$G_c$ (J/m$^2$)}
  & 8.7  & liquid N$_2$ & DCC          & Wiederhorn               & \cite{wiederhorn1969fracture} \\
  & 7.3  & --           & theoretical  & Rouxel \& Yoshida        & \cite{rouxel2017_1} \\
  & 5.2\textsuperscript{$\dagger$} & vacuum & CNMB (bowtie) & Cui \& Vinci & \cite{cui2017chevron} \\
\midrule
\multirow{2}{*}{$\ell_c$ (nm)}
  & 3.0  & --           & PF (theoretical)  & Tomić \textit{et al.}    & \cite{tomic2022phase} \\
  & 2.0  & --           & MD           & Molnár \& Barthel        & \cite{molnar2024_1} \\
\midrule
\multirow{8}{*}{$K_{IC}$ (MPa$\cdot$m$^{1/2}$)}
  & 0.73 & dry air      & NB & Smith \& Chowdary        & \cite{smith1975fracture} \\
  & 0.78 & air          & CNSR         & Lucas \textit{et al.}    & \cite{lucas1995_1} \\
  & 0.65 & vacuum       & CNMB         & Mueller \textit{et al.}  & \cite{mueller2015fracture} \\
  & 0.67 & vacuum       & CNMB         & Žagar \textit{et al.}    & \cite{zagar2016fracture} \\
  & 0.58 & air          & NI  & Harding \textit{et al.}  & \cite{harding1994_1} \\
  & 0.68 & vacuum       & PS & Bruns \textit{et al.}    & \cite{bruns2020fracture} \\
  & 0.60 & air          & NI  & Scholz \textit{et al.}   & \cite{scholz2004fracture} \\
  & 0.71 & vacuum       & NMB & Liu \& Shinozaki         & \cite{liu2023_1} \\
\midrule
\multirow{7}{*}{$\sigma_c$ (GPa)}
  & 5.9  & air          & $\mu$-fiber tension & Proctor \textit{et al.} & \cite{proctor1967strength} \\
  & 6.0  & air          & 2PB          & Baikova \textit{et al.}  & \cite{baikova2013method} \\
  & 5.1  & air          & 2PB          & Tang \textit{et al.}     & \cite{tang2015_1} \\
  & 0.14 & air          & ROR          & To \textit{et al.}       & \cite{to2018_1} \\
  & 11.5 & liquid N$_2$ & 2PB          & Tang \textit{et al.}     & \cite{tang2015_1} \\
  & $\sim$10 & air   & nano-fiber tension & Brambilla \& Payne   & \cite{brambilla2009_1} \\
  & 10--13\textsuperscript{$\ddagger$} & vacuum (TEM) & nano-fiber tension & Luo \textit{et al.} & \cite{luo2016_1} \\
\bottomrule
\end{tabular}
\end{table}

In the next section we will discuss the role of the beam geometry on the fracture response and the effect of measurement error on the estimated $G_c$ values. Finally, the connection between brittle properties and atomic structure of silica will be discussed.

\section{Discussion}\label{sec:discussion}

In the previous section, the brittle properties of silica glass were successfully evaluated through the combined microbeam test and phase-field simulations. The results were consistent with previous literature and provide guidelines for the evaluation of the brittle behavior of other amorphous materials. In this section we discuss the following topics: a) the role of microbeam geometry (notch depth $h_s$ and crack tip diameter $D$) on the fracture response (fracture force), b) the effect of error in geometry measurements on the estimated $G_c$ values, and finally c) the connection between brittle properties and structure of silica glass and possible implications on silicate glasses.

\subsection{Role of microbeam's geometry on fracture response}

We performed a parametric study on notched beams by changing the notch geometry (depth and tip diameter) and material toughness ($G_c$). This analysis provides information on how the geometry affects the fracture response (brittle vs. strength regime).

\begin{figure}[H]
    \centering
    \includegraphics[width=0.6\columnwidth]{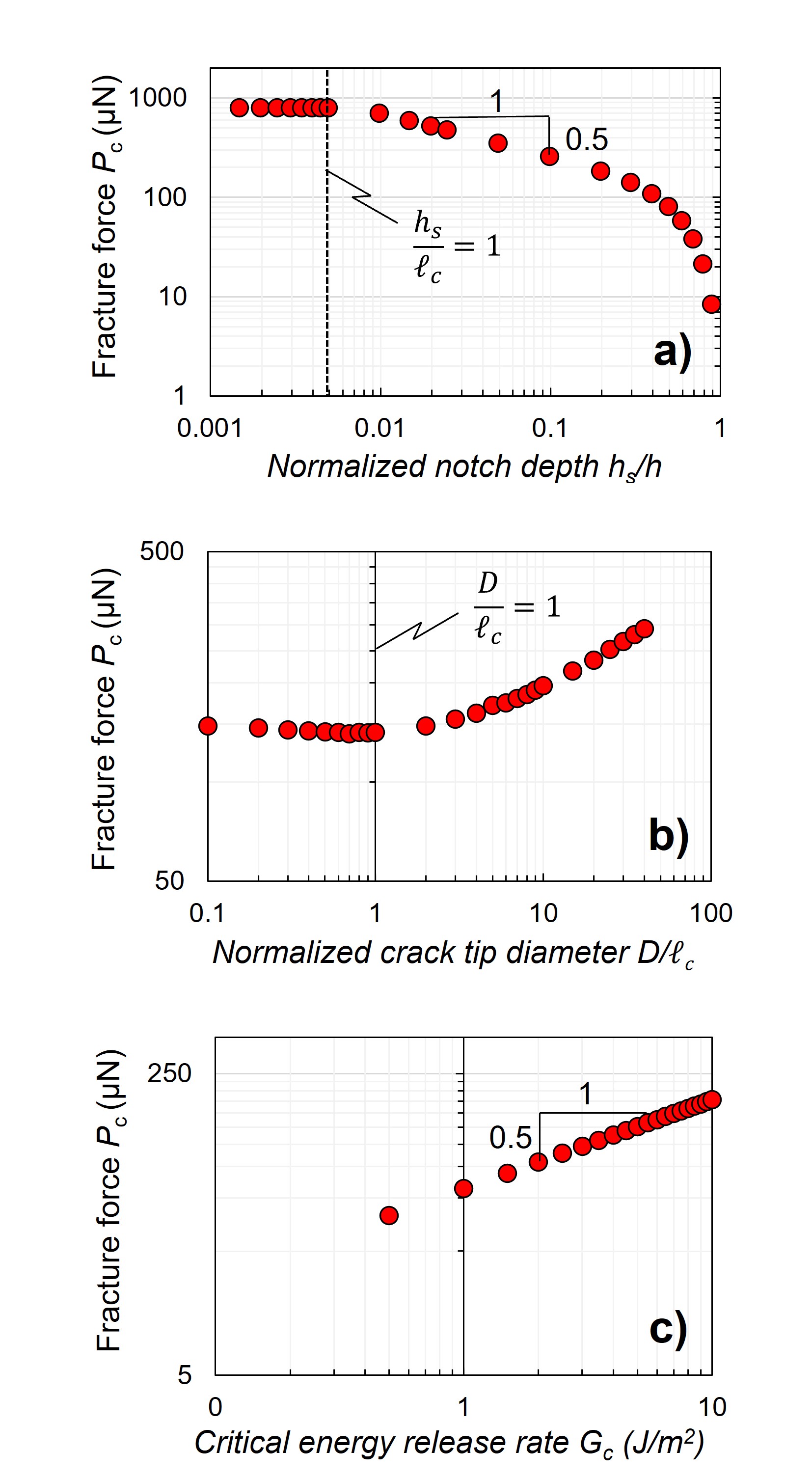}
    \caption{ Effects of microbeam geometry on the observed notched beam fracture forces $P_c$ (A) effect of notch depth normalized by the total beam height $h$, (B) effect of the notch diameter normalized by the length-scale parameter $\ell_c$, and (C) effect of the critical energy release rate.}
    \label{fig:effects}
\end{figure}

Figure \ref{fig:effects}A shows the fracture force as a function of the notch depth normalized by the total height of the beam. For very small notch depths ($h_s/\ell_c<1$) having a crack tip radius smaller than $\ell_c$, the fracture force is independent of the notch depth; this corresponds to the strength regime, in which the material fails upon reaching its tensile strength. When the depth of the notch goes beyond $h_s/\ell_c=1$, the fracture force decreases with notch size at a rate of 0.5 in log scale, corresponding to the Griffith brittle fracture regime. This regime remains constant until a $h_s/h$ ratio of about 0.4. At larger depths, the relationship between the fracture force and depth becomes highly non-linear due to the boundary effects (bottom side of the beam); in this region the stiffness of the beam decreases rapidly with notch depth. In our current investigation, the values of $h_s/h$ were between 0.25 and 0.5, therefore near the boundary between the Griffith and non-linear brittle regimes. 
The equation proposed by Matoy \textit{et al.} for the evaluation of $K_{IC}$ (eq.~\ref{eqn:Matoy}) using geometrical considerations of the microbeam is valid for $h_s/h$ values between 0.05 and 0.45. The values of $K_{IC}$ obtained from $G_c$ through phase-field calculations (plane strain approximation) are valid only as long as we are in the brittle Griffith regime. 
\\
The relationship between the fracture force and notch diameter (normalized by $\ell_c$) at constant notch depth $h_s$ is shown in Figure \ref{fig:effects}B. The fracture force $P_c$ is constant as long as the diameter $D$ is smaller than $\ell_c$. It is noted that the fracture force slightly increases when the tip radius approaches zero as assumed in Linear Elastic Fracture Mechanics (LEFM). When the notch diameter goes beyond $\ell_c$, the fracture force increases and the material failure cannot be described by Griffith fracture. Finally, the fracture forces for a constant, sufficiently large notch depth ($h_s$) follow a linear relationship with $G_c$ in log scale with a slope of 0.5, in accordance with Griffith's law (Figure \ref{fig:effects}C).

These results provide guidelines for the geometrical design of the microbeams. In this sense, by controlling the geometry we can predict whether the microbeam will behave as a Griffith brittle solid or will fracture through strength-controlled failure.

\subsection{Effect of geometry measurement error on evaluation of $G_c$}
Errors in the geometrical measurement of the cantilevers can result in significant errors in the evaluation of their brittle properties ($G_c$ and $K_{IC}$) and stiffness. Mueller \textit{et al.} evaluated the effect of error in geometry measurement on the resulting error in stiffness and fracture toughness of Chevron notched microbeams \cite{mueller2015fracture}. Their analysis suggests that the error in the measurement of the height ($h$) as well as the depth of the notches has the most significant effect on the evaluated stiffness and fracture toughness. In their work, an error of 5\% in those geometrical parameters can result in a 10--30\% error in estimated stiffness and fracture toughness. This behavior is expected since the stiffness of microbeams with fixed cross sections is proportional to the cube of their height ($\propto h^3$) and to their width ($\propto w$). We conducted a similar analysis on the investigated microbeam geometries. The results are summarized in table \ref{table:sensitivity}. For the notched beams, the most significant effect on the evaluation of $G_c$ and stiffness originates from the error in the measurement of the microbeam height $h$, followed by its length $L$ and width $w$. For the bone beam, the same trends are observed; however, the height has less influence in the case of the bone geometry compared to the notched microbeams. We define here the stiffness as the elastic slope of the load-displacement curve during the bending of the beams as evaluated by the FEA calculations. It is noted that for the bone beams, error in the geometry measurements of the narrow section can also have a strong impact on the calculated $G_c$, particularly the parameter $b$. The effect of error in the elastic modulus remains proportional in the evaluation of both $G_c$ and stiffness for both geometries.

\begin{table}[H]
\centering
\caption{Effect of 5\% increment change of the geometrical parameters ($L$, $w$, $h$, $\ell_s$, $h_s$, $\ell_p$, $\delta$, $a$, $b$) and Young's modulus $E$ on the percentage (\%) variation in estimated critical energy release rate $G_c$ and elastic stiffness (values in parentheses). Reference geometrical parameters and fracture forces have been used for each microbeam type.}
\label{table:sensitivity}
\vspace{1em}
\setlength{\tabcolsep}{4pt}
\small
\begin{tabular}{ccccccccccc}
\hline
\textbf{Type} & \textbf{$L$} & \textbf{$w$} & \textbf{$h$} & \textbf{$\ell_s$} & \textbf{$h_s$} & \textbf{$\ell_p$} & \textbf{$\delta$} & \textbf{$a$} & \textbf{$b$} & \textbf{$E$} \\
\hline
Notched & 19 (-15) & -9 (4) & -45 (50) & -2 (1) & 6 (-2) & -4 (3) & - & - & - & -4 (5) \\
Bone    & 17 (-14) & -15 (6) & -15 (12) & -2 (1) & - & -4 (4) & -1 (1) & -5 (-1) & 19 (-2) & -5 (5) \\
\hline
\end{tabular}
\end{table}

\subsection{Brittle properties of amorphous silica glass and connection with its atomic structure}
Our evaluations of the brittle properties of silica glass reveal that it is a material with multi-GPa strength and a damage regularization size ($\ell_c$) of a few nanometers. This length-scale goes beyond the local structure (coordination polyhedra) and medium range order (polyhedra connectivity) of silica glass. Recent work by Molnár and Barthel \cite{molnar2024_1} showed through atomic-scale simulations of silica glass that the fracture surface energy $\gamma$ comprises a contribution from the free surface energy and a major contribution (four times larger) from damage, defined as the local reduction in the material's stiffness without plasticity. The diffuse damage could extend over 2 nm from the crack surface inside the material, correlating spatially with the number of broken rings in the structure. In addition, they confirmed that the $\ell_c$ obtained from the homogeneous phase-field solution was close to that obtained through MD calculations. Our results show that the diffuse damage regions is within a few nanometers similar to that found in MD simulations. Although MD simulation results are strongly dependent on the choice of semi-empirical potential, our results support that $\ell_c$ is not just a numerical parameter but connected to the material's atomic structure.

The effect of tensile stress on the fracture response and atomic structure of silicate glasses has been evaluated in the past. Taniguchi and Ito~\cite{taniguchi2008_1} showed by molecular dynamics simulations of a soda-lime-silicate glass under uniaxial tension that the structure exhibits three main changes: expansion of T-O-T angles, changes in ring statistics (small rings form at the expense of larger rings), and growth of cavities. Tamura \textit{et al.} \cite{tamura2011_1} evaluated the uniaxial tension behavior of silica glass from first-principles calculations where they found that rotations of SiO$_4$ tetrahedra serve as the main stress release mechanism during elastic tension. Pedone \textit{et al.} \cite{pedone2008_1} investigated the structural changes of silica under tension under different boundary conditions. During the straining process, both the Si-O distances and Si-O-Si angles increase linearly until the failure point. In parallel, the ring distribution becomes broader, enhancing the formation of both small and large rings at the expense of medium-size rings, followed finally by void coalescence and fracture. It was noted that crystalline SiO$_2$ (cristobalite) has an intrinsic strength of about three times that of its glass counterpart, attributed to its single-size ring distribution compared to silica glass, suggesting a smaller length-scale. As shown by Moln{\'a}r and Barthel~\cite{molnar2024_1}, the ring changes seem to relate closely to the damage of the material. Du and Cormack \cite{du2004_1} showed in sodium silicate glasses that increasing the amount of sodium at the expense of silica promotes the formation of large-membered rings at the expense of the main 6-membered rings observed in silica glass. Therefore, in glasses with a large amount of non-bridging oxygens or network modifiers and broad ring distributions, damage is expected to be more diffuse than in regularly-sized ring structures such as those of crystalline SiO$_2$ or silica glass.

Since $\ell_c$ governs the spatial extent over which the elastic energy is dissipated through bond rupture resulting in stiffness reduction, materials whose networks accommodate stress through the rearrangement of a broader distribution of ring sizes are expected to exhibit a more delocalized (diffuse) damage zone, and therefore a larger $\ell_c$. Following equation \ref{eqn:strength}, where the strength is inversely proportional to $\ell_c$ and proportional to $G_c$, this larger length-scale translates directly into a reduced intrinsic strength $\sigma_c$. Amorphous silica, with its comparatively narrow ring-size distribution and highly connected tetrahedral network, thus represents one limiting case in oxide glasses of a well-regularized structure with a small $\ell_c$ and consequently a high intrinsic strength. It is important to note that the fracture surface energy $\gamma$ contributes to the ultimate strength of the material through $G_c$. For the particular case of silica, both the Si-O bond density and the high bond energy contribute to the large $G_c$ of amorphous silica \cite{rouxel2017_1,rouxel2017_2}.

This structural interpretation, although consistent with our observations, remains qualitative at present. It therefore remains for future work to evaluate, for oxide glasses, the effect of different network formers (NWF) such as boron oxide (B$_2$O$_3$), phosphorus oxide (P$_2$O$_5$), or aluminum oxide (Al$_2$O$_3$), as well as the role of network modifier (NWM) type and non-bridging oxygens, on the length-scale $\ell_c$ of the glass, helping to establish a more accurate description of brittle fracture and strength across glass families. Such a framework would ultimately allow the brittle response of a glass to be anticipated from its composition and short- to medium-range order, providing a rational basis for the design of stronger and more damage-tolerant glass compositions. Finally, although amorphous silica was completely brittle in tension, the partial existence of plasticity under tension in other silicate glasses or brittle materials remains possible.

\newpage

\section{Summary and Conclusions}\label{sec:conclus}

In this work we presented a new approach for the evaluation of the brittle properties of amorphous silica: the critical energy release rate $G_c$, the intrinsic length-scale $\ell_c$, the stress intensity factor $K_{IC}$ and the ultimate tensile strength $\sigma_c$. The approach combines microbeam bending experiments with phase-field fracture calculations. The phase-field method uses a thermodynamically consistent formulation based on the direct transformation of elastic energy to fracture surfaces enabling tracking of crack nucleation and propagation from singularities without requiring an explicit fracture surface to be prescribed \textit{a priori}. The main conclusions are summarized as follows:

\begin{itemize}
\item Two complementary FIB-fabricated geometries were tested on amorphous silica, each probing a different fracture regime. Single-notched microbeams, whose sharp singularities emulate the Griffith brittle regime, allowed the determination of $G_c$, describing the material's toughness, independently of $\ell_c$. The newly designed bone-shaped microbeams with blunt singularities gave access to the strength-controlled regime, enabling the evaluation of the material's intrinsic length $\ell_c$ and, therefore, of the ultimate tensile strength $\sigma_c$.This approach extends brittle-fracture characterization, beyond toughness-only approaches commonly used in micromechanical studies.
\item The intersection of the fracture-force-consistent $G_c$--$\ell_c$ curves from both geometries yielded, for amorphous silica, $G_c = 5.1$~J/m$^2$, $K_{IC} = 0.61$~MPa$\cdot$m$^{1/2}$, $\ell_c = 9.1$~nm, and $\sigma_c = 6.8$~GPa. These values agree closely with previously reported data. To our knowledge, this is the first experimentally derived regularization length-scale for amorphous silica.
\item A parametric study clarified how the notch depth and tip radius govern the transition between the strength-controlled and Griffith brittle regimes, providing practical guidelines for the geometrical design of microbeam experiments.
\item In agreement with previous works by Mueller \textit{et al.}~\cite{mueller2015fracture}, errors in dimensional measurement (particularly of the beam height $h$ and notch depth $h_s$) translate into substantial uncertainty in the determined $G_c$ and stiffness,  pointing out the impact of geometric characterization.
\item Beyond amorphous silica, the proposed methodology offers a general framework for simultaneous evaluation of toughness and strength in brittle solids, including metallic glasses, polymers, ceramics, or other amorphous systems. 

\end{itemize}


\section*{CRediT authorship contribution statement}
\textbf{Gustavo Alberto Rosales-Sosa}:  Writing - original draft, Conceptualization, Formal Analysis, Data curation, Methodology, Project administration, Software, Investigation, Validation, Visualization.
\textbf{Gergely Molnár}: Writing - review \& editing, Conceptualization, Methodology, Software, Supervision.
\textbf{Yoshinari Kato}: Writing - review \& editing, Conceptualization, Methodology, Investigation, Supervision, Project administration.
\textbf{Taichi Nakashima}: Writing - review \& editing, Investigation, Methodology.
\textbf{Takahito Ohmura}: Writing - review \& editing, Supervision, Methodology, Resources.
\textbf{Etienne Barthel}: Writing - review \& editing, Supervision, Methodology.
\textbf{Guillaume Kermouche}: Writing - review \& editing, Supervision, Methodology.
\textbf{Sergio Sao Joao}: Writing - review \& editing, Methodology.
\textbf{Shingo Nakane}: Writing - review \& editing, Supervision, Funding Acquisition, Project administration.
\textbf{Hiroki Yamazaki}: Supervision, Funding Acquisition, Project administration.

\section*{Declaration of competing interest}
The authors declare that they have no known competing financial
interests or personal relationships that could have appeared to influence
the work reported in this paper.

\section*{Acknowledgements}
This work was funded by Nippon Electric Glass Co., Ltd (NEG). The authors would like to thank Dr. Naoki Toyofuku and Mayu Yabe at NEG for all the valuable advice related to the FIB fabrication of the microbeams.  


\bibliography{references_mnd}

\end{document}